\documentclass[aps,eqsecnum]{revtex4}
\usepackage{epsfig,rotating,amsmath,amssymb,graphics,rotate,psfrag}

\begin{document}
\tableofcontents
\section{Highlights from the presentations}
%
%
\subsection{K. Blaum, A. Doerr, C. E. Duellmann, K. Eberhardt, S. Eliseev, C. Enss, A. Faessler, A. Fleischmann, L. Gastaldo$^*$, S. Kempf, M. Krivoruchenko, S. Lahiri, M. Maiti, Yu. N. Novikov, P. C.-O. Ranitzsch, F. Simkovic, Z. Szusc, M. Wegner  : ``The Electron Capture $^{163}$Ho Experiment ECHo''}
\begin{description}
\it\small
\setlength{\parskip}{-1mm}
\item[\small\it K.B., A.D., S.E.:]Max-Planck Institute for Nuclear Physics Heidelberg, Germany
\item[\small\it C.E.D.:]Institute for Nuclear Chemistry, Johannes Gutenberg University, Mainz, Germany, GSI Helmholtzzentrum für Schwerionenforschung, Germany, Helmholtz Institute Mainz, Germany
\item[\small\it K.E.:]Institute for Nuclear Chemistry, Johannes Gutenberg University, Mainz, Germany, Helmholtz Institute Mainz, Germany
\item[\small\it C.E., A.F., L.G., S.K., P.C.O.R., M.W.:]Kirchhoff Institute for Physics, Heidelberg University, Germany
\item[\small\it A.F.:] Institute for Theoretical Physics, University of Tuebingen, Germany
\item[\small\it M.K.:]Institute for Theoretical and Experimental Physics Moscow, Russia
\item[\small\it S.L.:]Saha Institute of Nuclear Physics, Kolkata, India
\item[\small\it M.M.:]Department of Physics, Indian Institute of Technology Roorkee, India
\item[\small\it Yu.N.N.:]Petersburg Nuclear Physics Institute, Russia
\item[\small\it F.S.:]Department of Nuclear Physics, Comenius University, Bratislava, Slovakia
\item[\small\it Z.S.:]Institute of Nuclear Research of the Hungarian Academy of Sciences
\item[$^*$] Corresponding author, Loredana.Gastaldo@kip.uni-heidelberg.de
\end{description}
%

The Electron Capture $^{163}$Ho Experiment, ECHo, has the aim to investigate the electron neutrino mass in the energy range below $1\,$eV by a high precision and high statistics calorimetric measurement of the $^{163}$Ho electron capture (EC) spectrum.  $^{163}$Ho decays by capturing an electron from the inner atomic shells to an excited state of the $^{163}$Dy atom with a half-life of $\tau_{1/2}\approx 4570$ years and a recommended value for the energy available to the decay $Q_{\mathrm{EC}}\approx \, 2.5\,$keV which still has a large uncertainty since different experiments have measured values between 2.3 and 2.8 keV, as discussed in Section \ref{QEC}. The atomic de-excitation is a complex process which includes cascades of both x-rays and electron emissions (Auger electrons and Coster-Kronig transitions). By performing a calorimetric measurement of the de-excitation spectrum, i.e. by measuring for each event the sum of the energies of all emitted photons and electrons with the same detector, the sensitivity to the neutrino mass is increased. In order to perform a calorimetric measurement of the $^{163}$Ho EC spectrum, the $^{163}$Ho source has to be: i) part of the sensitive volume of detector in order not to have energy losses in the source that are not measured, ii) homogeneously distributed so that the detector response has not position dependent effects and iii) completely contained in the detector in order to ensure a quantum efficiency for the emitted particles of $100\%$.

In [1], several scenarios to reach the sub-eV sensitivity for the neutrino mass by the analysis of the calorimetrically measured $^{163}$Ho EC spectrum have been described. According to this work, the performance required for the detectors to measure the $^{163}$Ho EC spectrum are really demanding: first of all it should be possible to prepare the detector with the $^{163}$Ho source fully contained in the sensitive volume, then all the particles emitted in the $^{163}$Ho EC need to be detected with the same efficiency, an energy resolution $\Delta E_{\mathrm{FWHM}}$ better then 10 eV is asked and the signal rise-time $\tau_{\mathrm{r}}$ has to be short, possibly below $1\,\mu$s. Moreover in order to reach the aimed sensitivity, a statistics of $10^{14}-10^{16}$ counts in the full spectrum need to be acquired, depending on different combinations of detector parameters and on the defined value of $Q_{\mathrm{EC}}$. The measurement of the full statistics within a reasonable time of few years requires a total $^{163}$Ho activity of more than 1 MBq. The production of the $^{163}$Ho source with the needed activity and purity is a very important aspect of the ECHo project. In fact it is required that the presence of long living radioactive contaminants is negligible so that their contribution to the background of the $^{163}$Ho EC spectrum is also negligible and the presence of material from the target that has been used to produce the $^{163}$Ho atoms has to be about $10^{-2}\,-\,10^{-1}$ times the amount of $^{163}$Ho.

Once the high purity $^{163}$Ho source will be available and it will be embedded in the energy absorbers of detectors able to measure with high precision the $^{163}$Ho EC spectrum, it will be possible to perform the high statistics calorimetric measurement of the $^{163}$Ho EC spectrum. In order to extract a limit on the neutrino mass in the sub-eV range from this measurement, it is important to improve the knowledge of the expected $^{163}$Ho EC spectrum. A large part of the ECHo collaboration is working to determine theoretically and experimentally the parameters describing the $^{163}$Ho EC spectrum. In particular one of the important goals of the ECHo experiment is to define the energy $Q_{\mathrm{EC}}$ available to the $^{163}$Ho decay as the mass difference between $^{163}$Ho and $^{163}$Dy within an uncertainty of 1 eV. This value will be used as a reference point to investigate the high energy part of the $^{163}$Ho EC spectrum. Moreover it is important to quantify the modifications to the $^{163}$Ho EC spectrum due to solid state effects generated by the fact that the $^{163}$Ho ions are embedded in a solid.

\begin{figure*}
  \includegraphics[angle=0, width=.50\textwidth]{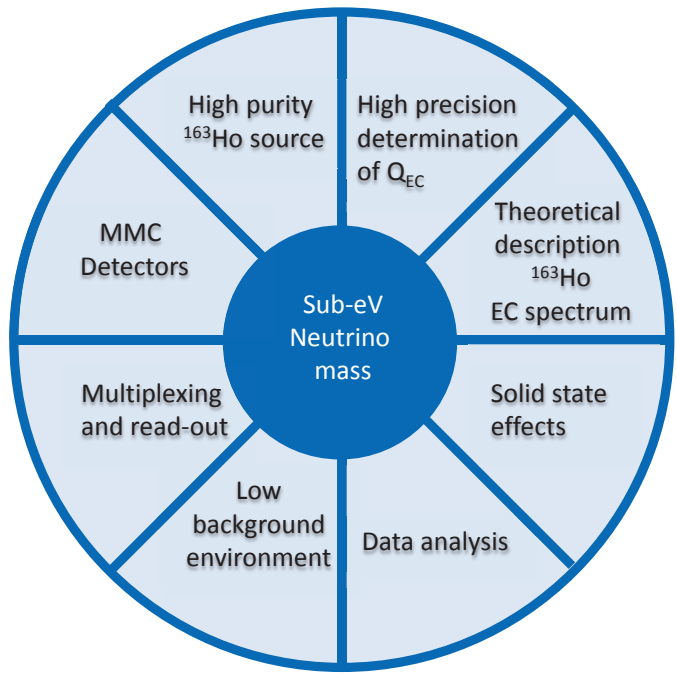}
  \caption{ Diagram showing the structure of the ECHo experiment.  \vspace{-5mm}}
  \label{ECHo_structure}
\end{figure*}

Fig. \ref{ECHo_structure} shows a diagram representing all the investigation routes which compose the ECHo experiment. In the following a few aspects of the ECHo experiment will be described. First of all the detector technology that will be used in ECHo will be introduced as well as the proposed read-out scheme. The results achieved by the first detector prototype will be also discussed. The challenges of producing and purifying the high activity and high purity $^{163}$Ho source will be presented and finally the high precision measurement of the $Q_{\mathrm{EC}}$ value will be described.

\subsubsection{Low temperature metallic magnetic calorimeters and microwave multiplexing}
Presently the only detectors that can measure energy below 3 keV with high precision are low temperature micro-calorimeters [2]. They make use of the calorimetric
principle where the absorption of energy produces an increase
of the detector temperature $\Delta T$ proportional to the deposited energy $\Delta E$ and to the
inverse of the detector heat capacity $C_{\mathrm{tot}}$. These detectors work at temperatures below 100 mK where the phononic and electronic contributions to the heat capacity is smallest. A small heat capacity is one of the key parameters to have a good signal to noise ratio. For detectors developed to measure soft x-rays, the heat capacity is of the order of $1\,$pJ/K or even smaller. It is then clear that only a small fraction of the activity of $^{163}$Ho, required to reach the sub-eV sensitivity to the neutrino mass, can be embedded in a single detector without degrading its performance by increasing its heat capacity. Moreover the reduced activity per detector is also important to reduce the so-called un-resolved pile-up problem. The un-resolved pile-up consists in the impossibility to resolve two or more events happening in the same detector within a time interval of less than the signal rise-time. In this case the detector would show a single event, with an energy given approximately by the sum of the energy of the single events, which contributes to the background. In first approximation the fraction of un-resolved pile-up events is given by the product of the total activity in the detector and the pulse rise-time. Therefore in order to reduce these un-wanted events a fast detector response and a reduced activity per pixel are important features of the detector. On the other hand the possibility to increase the activity per pixel will reduce the number of detectors needed to perform the experiment. It is therefore important to optimize the activity per pixel in order on one hand  to have high energy resolution and low intrinsic background and on the other hand to reduce the number of pixels.

As it has been discussed in the previous section, the performance required to the detectors which allows for the achievement of the sub-eV sensitivity to the electron neutrino mass are: high precision detection of the energy of electrons and photons below 3 keV which corresponds to an energy resolution $\Delta E_{\mathrm{FWHM}}\,\leq\,10\,$eV, a fast signal rise-time $\tau_{\mathrm{r}}\,\leq\,1\, \mu$s and a read-out scheme that makes possible a fast and high pixels density multiplexing system.

Within the ECHo experiment low temperature metallic magnetic calorimeters (MMCs) [3] will be used. MMCs are energy dispersive
detectors typically operated at temperatures below $50\,$mK. In first approximation these detectors consist of a particle absorber, where the energy is deposited, tightly connected to a temperature sensor which is then weakly connected to a thermal bath. As for all the other micro-calorimeters, also for MMCs the deposition of energy in the absorber leads to an increase of the detector temperature. The temperature sensor of the MMCs is a paramagnetic alloy which resides in a small magnetic field. The change of temperature leads to a
change of magnetization of the sensor which is read-out as
a change of flux by a low-noise SQUID magnetometer. The
sensor material, presently used for MMCs, is a dilute alloy of erbium in gold,
Au:Er. The concentration of erbium ions in the sensor can be chosen to
optimize the detector performance and usually varies between 200
ppm and 800 ppm.
The spectral resolving power of a state of the art MMCs
for soft x-rays is above 2000. For completely micro-structured detectors, an energy resolution of $\Delta E_{\mathrm{FWHM}}\,=\,2\,$eV at $6\,$keV and a signal rise-time $\tau_{\mathrm{r}}\,=\,0.09\, \mu$s [4] have been achieved. Moreover the read-out scheme for MMCs is compatible with several multiplexing techniques developed for low temperature micro-calorimeters, in particular with the microwave multiplexing as will be discussed in the following. The achieved performance suggests that MMCs are suitable detectors for measuring the high precision and high statistics EC spectrum of $^{163}$Ho.

\begin{figure*}
  \includegraphics[angle=0, width=.80\textwidth]{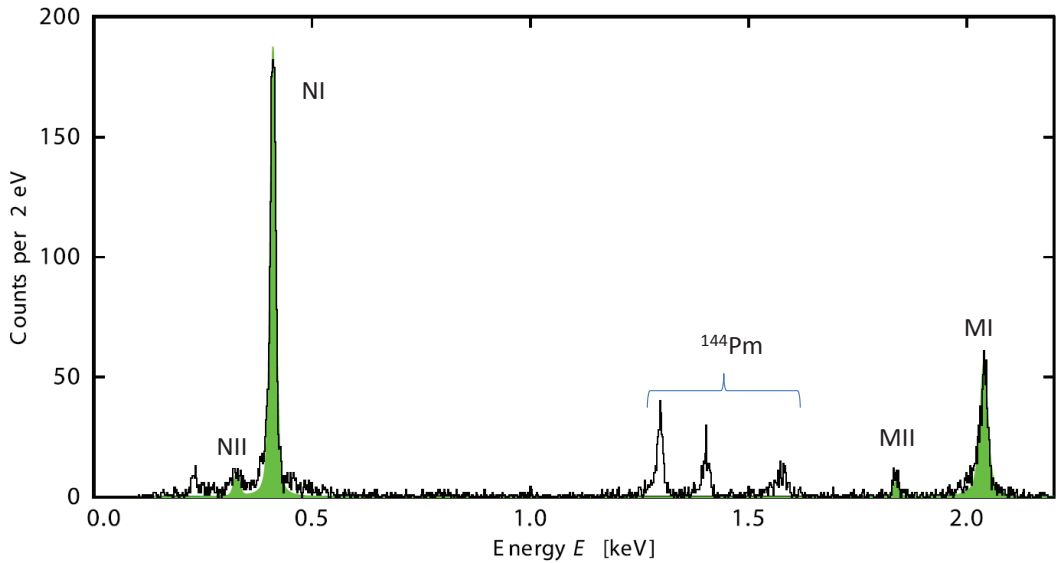}
  \caption{ Calorimetric EC spectrum of $^{163}$Ho as measured (black histogram) and fitted (green
area) The calorimetric lines MI, MII, NI, NII of the $^{163}$Ho EC spectrum can be perfectly seen. The additional lines present in the measured spectrum derive from the EC decay of $^{144}$Pm which was implanted as PmF$^+$ ionized molecules together with the $^{163}$Ho ions. Reprinted from [7]  \vspace{-5mm}}
  \label{163Hospectrum}
\end{figure*}

A first test experiment to investigate the behavior of a MMC detector with $^{163}$Ho ion-implanted in the absorber has been successfully performed [5]. The $^{163}$Ho was produced at ISOLDE-CERN [6] by irradiating with a proton beam a Ta-W target. After surface ionization, a mass-selected beam with ions having mass $163\,$u was directed onto the detector chip and collimated on a surface having the diameter of about $2\,$mm.
\begin{figure*}
  \includegraphics[angle=0, width=.95\textwidth]{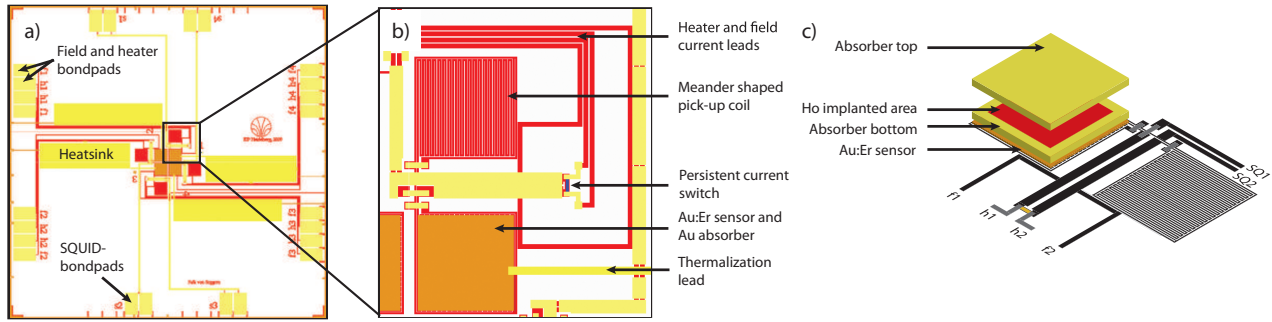}
  \caption{a) Schematic of a MMC detector chip for the $^{163}$Ho implantation experiment. The four double-meander pick up coils are located in the center of the chip. b) The magnified region shows a detailed picture of one detector. Only one side of the two double-meander pick-up coil is equipped with sensor and absorber. c) A simplified three-dimensional picture of one detector is shown. In red is indicated the position where the $^{163}$Ho ions are implanted. Reprinted from [7]  \vspace{-5mm}}
  \label{Det_scheme}
\end{figure*}
Fig. \ref{Det_scheme}a) shows the layout of the first prototype of detector chip for the measurement of the $^{163}$Ho EC spectrum. The chip is equipped with four detectors. The detectors are based on the niobium double-meander pick-up coil geometry [3]. The details of the single detector are shown in the magnification, Fig. \ref{Det_scheme}b). In particular only one side of the double-meander pick-up coil has been equipped with Au:Er sensor and absorber to better characterize the thermo-dynamical properties of sensor and absorber materials. The description of the chip design and fabrication is given in [5]. A schematic cross-section of the detector is shown in Fig. \ref{Det_scheme}c).
The absorber is composed of two gold layers, each of dimensions $190\,\times\,190\,\times\,5\,\mu$m$^3$. On top of the first gold layer, indicated as "absorber bottom", the area where the $^{163}$Ho is implanted is indicated in red and has dimensions $160\times160\,\mu$m$^2$. This area is smaller than the absorber area in order to avoid loss of energy through the side walls of the absorber and therefore to achieve the complete quantum efficiency. During the implantation process, the complete chip besides the four $160\times160\,\mu$m$^2$ squares was protected with photo-resist.
The $^{163}$Ho activity in each pixel was approximately $10^{-2}$ Bq, corresponding to about $10^{10}$ implanted ions. A few tests have been performed with these detectors showing that the implantation process did not degrade the performance of the MMC [5]. With this first measurements, the presently most precise calorimetric measurement of the $^{163}$Ho EC spectrum was obtained. Fig. \ref{163Hospectrum} shows the measured spectrum. The largest background is due to the EC of $^{144}$Pm that was mass-selected and implanted as PmF$^+$ together with the $^{163}$Ho ions. Presently many efforts are dedicated to the production of high purity $^{163}$Ho sources as will be discussed in the next section. The detailed description of this $^{163}$Ho EC spectrum is discussed in [7]. The measured energy resolution was $\Delta E\,\simeq\,12\,$eV and the rise-time $\tau_{\mathrm{r}}\,\simeq\,100\,$ns. The position of the energy peaks was defined within few eV and the obtained best estimation of the total energy available to the $^{163}$Ho decay was $Q_{\mathrm{EC}}\,=\,(2.80\,\pm\,0.08)\,$keV.

The results achieved with the first prototype of MMC featuring an absorber with implanted $^{163}$Ho ions and the possibility to improve the energy resolution of the detectors, as discussed in [5] indicate that MMCs meet all the requirements to be used for the high precision measurement of the $^{163}$Ho EC spectrum to investigate the electron neutrino mass. The aim of the ECHo collaboration is to develop MMCs with $10^{11}-10^{13}$ $^{163}$Ho ion in each absorber, corresponding to an activity per pixel of 1 to 100 Bq, having an energy resolution below 3 eV and a pulse rise-time below 100 ns.
A very important aspect of the optimization of the MMCs is the production of the absorber containing the $^{163}$Ho ions. In particular different materials with different concentrations of $^{163}$Ho will be tested as well as different methods to insert the $^{163}$Ho ions in the energy absorbers of MMCs will be investigated as for example ion-implantation and thermo-reduction of the $^{163}$Ho source followed by the preparation of an alloy.

In order to reach the $^{163}$Ho activity to perform the experiment the total number of detectors ranges from $10^4$ to $10^6$. The read-out of such a large number of MMCs can not be done by using the single channel readout since it would produce an enormous heat load on the experimental platform. Therefore it is important to develop a suitable method to multiplex MMCs detectors. The requirements that this method has to fulfill are: low dissipation on the experimental platform, relatively large bandwidth for each pixel of the array and 
ideally no additional noise compared to single pixel read-out.
All these requirements are met by the microwave SQUID multiplexing [8].
In this multiplexing scheme every detector is coupled to a non-hysteretic, un-shunted rf-SQUID which is coupled to a superconducting microwave resonator with high internal quality factor and unique resonance frequency. A change of magnetic flux inside the SQUID caused by an event in the detector leads to a change of the effective SQUID inductance and therefore, due to the mutual interaction, to a change of the resonance frequency of the corresponding microwave resonator. It is possible to measure the signal of each detector simultaneously by capacitively coupling the corresponding number of resonators to a common transmission line, injecting a microwave frequency comb driving each resonator at resonance and monitoring either amplitude or phase of each frequency component of the transmitted signal.

Based on experimental results obtained with the first prototype SQUID multiplexer and numerical simulations, the current multiplexer design concerning rf-SQUID layout, SQUID-to-resonator coupling and fabrication of the Nb/Al-AlOx/Nb Josephson junctions [9] have been optimized. Presently new chips consisting of 64 pixels which are read-out using the microwave multiplexing scheme have been developed. After their characterization, these chips will be equipped with absorbers containing $^{163}$Ho and will be used for a small scale experiment.

\subsubsection{$^{163}$Ho source: production and purification}
The production of the proper amount of $^{163}$Ho atoms and the methods to purify this source from the target material and from other isotopes, which are produced within the same process, are of paramount importance for the success of the ECHo experiment. Preliminary studies of different approaches to produce $^{163}$Ho have already been performed by the ECHo collaboration. The production methods for $^{163}$Ho can mainly be divided into two branches: i) charged particle activation of suitable targets in a direct way, that is the optimization of the $^{163}$Ho production, or in an indirect way, that is the optimization of the production of the $^{163}$Ho precursor $^{163}$Er which has a half-life $T_{1/2} \,=\,75\,$min and ii) thermal neutron activation of enriched $^{162}$Er targets.
The reaction that is typically used for the direct activation with proton beams is $^{\mathrm{nat}}$Dy$(p,xn)^{163}$Ho. Another possible direct reaction uses deuteron projectiles $^{163}$Dy$(d,2n)^{163}$Ho. Examples for an indirect way of $^{163}$Ho production are $^{\mathrm{nat}}$Dy$(\alpha,xn)^{163}$Er($\epsilon$)$^{163}$Ho and $^{159}$Tb$(^7$Li$,3n)^{163}$Er($\epsilon$)$^{163}$Ho.

In the ECHo experiment all the described methods for production of $^{163}$Ho through charged particle activation processes as well as with neutron irradiation of a $^{162}$Er target will be extensively investigated as well as possible new methods. Moreover methods for the separation of the $^{163}$Ho needs to be optimized in order to have in the final source only traces of the target material and in order to remove the radioactive contaminants to the level in which their contribution to the background of the calorimetric measurement is smaller then the intrinsic pile-up background. A few processes have already been tested. A liquid-liquid extraction technique to separate erbium nuclides from the dysprosium target was performed on $\alpha$-irradiated natural dysprosium target at Saha Institute for Nuclear Physics in Kolkata. The developed chemical method is fast enough to complete the entire chemical process within one half-life of $^{163}$Er. On a $^{162}$Er enriched erbium sample which was irradiated for 11 days with thermal neutrons at the BER II reactor at the Helmholtzzentrum in Berlin (the thermal neutron flux is $\Phi = 1.3 \times 10^{14}\,$s$^{-1}$cm$^{-2}$) an ion-chromatography process using $\alpha$-hydroxyisobutyric acid was used to separate the holmium ions from the erbium target ions. This source after the purification step contains about $10^{16}$ $^{163}$Ho ions and will partially be used for the characterization of the contaminants as well as for the optimization of the separation and purification method. Moreover a large fraction of this source will be used for future detector tests as well for first experiments to determine the $Q_{\mathrm{EC}}$ value with Penning Traps.

In the next future more tests will be performed both at accelerator facilities as well at reactor facilities. The aim of these tests is to investigate and quantify the production of radioactive contaminants and to improve the purification methods in order to reach the required purity.

\subsubsection{$Q_{\mathrm{EC}}$ determination}
\label{QEC}
The total energy available for the decay of $^{163}$Ho to $^{163}$Dy is a fundamental parameter to reach the sub-eV sensitivity to the electron neutrino mass by the analysis of the high energy part of the $^{163}$Ho EC spectrum. The recommended value is $Q_{\mathrm{EC}}\,=\,(2.55\,\pm\,0.016\,)$keV as can be found in [10], but other measurements give values that range from about 2.3 keV [11] to about 2.8 keV obtained by two calorimetric measurements performed using low temperature detectors [12] and [7]. It is to notice that the $Q_{\mathrm{EC}}$ value measured in the cited experiments, has been derived by the analysis of the EC spectrum of $^{163}$Ho.

A very strong method to determine the energy available to the $^{163}$Ho which is also independent from the decay process is the determination of the mass difference between mother and daughter atoms. Within the ECHo experiment the $Q_{\mathrm{EC}}$ of $^{163}$Ho will be measured by means of Penning traps [13].
The superiority of Penning-trap mass spectrometry over the other mass-measurement
methods lies in a determination of the mass $M$ of the nuclide of interest via the direct
measurement of the cyclotron frequency $\nu_{\mathrm{c}}$ of its ionic state with the electric charge $q$ in
a strong static homogeneous magnetic field $B$:
\begin{equation}
\nu_{\mathrm{c}}=\frac{1}{2\pi}\times \frac{q}{M} \times B
\end{equation}
Nevertheless, to perform such a measurement the ion must be confined to a well-localized
volume within the homogeneous magnetic field for at least some seconds. This is achieved by
a superposition of a static three-dimensional quadrupole electric field on the magnetic field,
such that an electrostatic potential well along the magnetic field lines is created. In such a
configuration of the fields, called the Penning trap, the magnetic field confines the motion of the ion
to the plane perpendicular to the magnetic field lines and the electrostatic field does not allow
the ion to escape along the magnetic field lines.
The presence of the electrostatic quadrupole field modifies the pure cyclotron motion of
the ion in the magnetic field to three independent trap motions: two radial motions which are the modified
cyclotron and the magnetron motions with the frequencies $\nu_-$ and $\nu_+$, respectively, and the axial
motion with the frequency $\nu_{\mathrm{z}}$. None of these frequencies are simple functions of the ion's
mass, but the sum of the radial frequencies is equal to the cyclotron frequency $\nu_{\mathrm{c}}$ on the level
of the required accuracy for traps employed in high-precision Penning-trap mass spectrometry.
Thus, a measurement of the radial eigenfrequencies results in a determination of the
cyclotron frequency of the ion. There are two methods to measure the cyclotron frequency. (1)
The cyclotron frequency is measured via detection of the image current induced by the ion's
motion in the resonant tank circuit attached to the trap [14]. (2) In the so-called time-of-flight
ion-cyclotron-resonance technique (ToF-ICR) the cyclotron frequency is determined from the
measurement of the time of flight of the ion through the strong gradient of the magnetic field [15].

The aim of the ECHo collaboration is to reach a precision on the $Q_{\mathrm{EC}}$ of 1 eV or better. This task is planned to be accomplished in the near future by the novel Penning-trap mass spectrometer PENTATRAP [16,17]. The uniqueness and complexity of this Penning-trap mass spectrometer is conditioned by the unprecedented relative accuracy of a few parts in $10^{12}$ with which the $Q_{\mathrm{EC}}$ must be measured. This requires a careful stabilization of the magnetic field, screening the magnet from stray electrical and magnetic fields and a temperature-stabilized experimental room with a vibration-free concrete cushion for the magnet. It is necessary to produce ions of heavy nuclides in very high charge states, which is done with an external electron-beam ion-trap source. Furthermore, the cyclotron frequencies of the decay mother and daughter nuclides have to be measured simultaneously. For this, five cylindrical Penning traps will be employed and the novel cyclotron-frequency measurement technique described in Ref. [18] will be applied.

\subsubsection{Conclusions}
The complexity of the ECHo experiment requires that the efforts of several working groups are combined to reach the aimed sensitivity in the sub-eV range for the electron neutrino mass. In the next future the first detector chip with integrated multiplexed read-out will be produced as well as new methods for the production and purification of the $^{163}$Ho source will be analyzed. 
The next goal of the ECHo collaboration is to perform an experiment of reduced size, about 100-1000 pixels read-out as few arrays. The uncertainty for the $Q_{\mathrm{EC}}$ value will be reduced to few tens of eV. With this first small scale experiment it will be possible to set an upper value on the neutrino mass in the few eV range which corresponds to an improvement of about two orders of magnitude compared to the present accepted limit of 225 eV [19].

\begin{center}
{\it References}
\end{center}
\begin{description}
\footnotesize	
\item[1] M. Galeazzi {\it et al.}, http://arxiv.org/abs/1202.4763
\item[2] C. Enss, Topics in Applied Physics, {\bf 99} (2005)
\item[3] A. Fleischmann {\it et al}, AIP Conf. Proc. vol. {\bf 1185} (2009) 571
\item[4] C. Pies {\it et al}, Journal of Low Temperature Physics {\bf 167} 3-4 (2012) 269
\item[5] L. Gastaldo {\it et al.}, Nuclear Inst. and Methods in Physics Research, A {\bf 711} (2013) 150
\item[6] E. Kugler, Hyperfine Interact. {\bf 129} (2000) 23
\item[7] P. C.-O. Ranitzsch {\it et al.}, Journal of Low Temperature Physics, {\bf 167} (2012) 1004
\item[8] J. A. B. Mates {\it et al.}, Applied Physics Letters {\bf 92} (2) [2008) 023514
\item[9] S. Kempf {\it et al.}, Supercond. Science And Technology {\bf 26} (2013) 065012
\item[10] A. Wapstra, G. Audi, C. Thibault, Nucl. Phys. A {\bf 729} (1), (2003) 129
          G. Audi, A. Wapstra, C. Thibault, Nucl. Phys. A {\bf 729} (1), (2003) 337

\item[11] J. U. Andersen {\it et al.}, Phys. Lett. B {\bf 113} (1982) 72
\item[12] F. Gatti {\it et al.}, Physics Letters B {\bf 398} (1997) 415
\item[13] K. Blaum,  Phys. Rep. {\bf 425} (2006) 1
\item[14] H.G. Dehmelt and F.L. Walls, Phys. Rev. Lett. {\bf 21} (1968) 3
\item[15] G. Graeff, H. Kalinowsky and J. Traut, Z. Phys. A {\bf 297} (1980) 35
\item[16] J. Repp {\it et al.}, Applied Physics B {\bf 107} (2012) 983
\item[17] C Roux {\it et al.}, Applied Physics B {\bf 107} (2012) 997
\item[18] S. Eliseev {\it et al.}, Physical Review Letters {\bf 110} (2013) 082501
\item[19] P. T. Springer {\it et al.}, Phys. Rev. A {\bf 35} (1987) 679

\end{description}
%
%
\end{document}